# Strategic Communication Protocols for Interstellar Objects Using a Threat-Communication Viability Index and the Information-Communication Paradox

David R. Gruber [1]

1  *D. R. Gruber, Dept. of Communication Studies, University of Nevada, Las Vegas*

Corresponding Address:

Department of Communication Studies

University of Nevada, Las Vegas

4504 S. Maryland Pkwy, GUA bld.

UNLV, Las Vegas, 89154

Email: David.Gruber@unlv.edu

Strategic Communication Protocols (SCPs) provide a structured approach for first contact with interstellar objects (ISOs) that demonstrate technological characteristics and high levels of threat. The SCPs find their starting point in an ISO Information-Communication Paradox, namely, as our knowledge of an ISO's threatening capabilities increases, the probability of successful communication decreases while the urgency of communication attempts simultaneously intensifies. From this paradox, a Threat-Communication Viability Index (TCVI) is created to describe when the value of communication attempts outweighs strategic silence. The TCVI scores the situation and operates as a decision-making tool for stakeholders tracking ISOs. The SCPs subsequently outline several diplomatic communication strategies in cases where the TCVI recommends communication.

# Introduction

Existing interstellar communication efforts inadequately address the unique challenges posed by interstellar objects (ISOs). METI International's messaging to distant star systems[1] and the SETI Permanent Committee's reply guidelines[2] both assume static, distant targets with extended communication windows. However, ISOs present fundamentally different circumstances. They are mobile, potentially threatening, and operate within compressed timelines that demand immediate decision-making.

The creation of the Loeb Scale for classification of ISOs based on threat (Eldadi, Tenenbaum, and Loeb, 2025) provides a foundation for communication protocols. The SCP framework does not assume communication is appropriate at every level of the scale. Instead, SCPs rely on a Threat-Communication Viability Index (TCVI) that considers the Loeb Scale in conjunction with time until ISO contact, available time frames for sending and receiving messages, as well as additional threat factors tied to hostility toward Earth and not within the Loeb Scale. That is, SCPs are designed for diplomatic communications, and communication attempts are reserved for circumstances with clear threats from technological ISOs. Choosing communication over silence in such cases is justified through five core propositions that define the context as well as nine base assumptions about ISOs that would make communication worth initiating.

To move forward, this paper first details the Information-Communication Paradox and then presents the five core propositions and nine assumptions about ISOs. After this, the paper describes the TCVI and how to use it. The final section outlines the SCP framework.

## I. The ISO Information-Communication Paradox

Central to the question of when and when not to communicate with a technological ISO is the following paradox: as an ISO displays increasingly sophisticated threat capabilities, the likelihood of achieving peaceful communication outcomes proportionally decreases. As threat indicators accumulate, communication becomes both more urgent and less likely to succeed.

---

[1] https://meti.org/mission
[2] https://iaaseti.org/en/protocols/

Proposing several specific threat indicators for an ISO 6 or above (ISO-6+) demonstrates how just one and certainly two increase the threat to Earth, injecting greater uncertainty, complexity, and fear into the communication situation. Although threat indicators are best developed by international teams, imagining one or more of the following being confirmed drives home the relationship between threat and the challenges of communication and hopes of success:

1. High-energy radiation emissions
2. Gravitational anomalies suggesting exotic physics applications
3. Time dilation or observed temporal displacement effects
4. Multiple coordinated objects
5. Abrupt course changes toward Earth
6. Sudden detection reduction or stealth behavior
7. Directed energy beams toward Earth
8. Swarm release or replication events
9. Communication jamming or interference with Earth systems

Confirmation of any of the above does not necessarily mean that an ISO would act against Earth, nor does it remove the possibility that an actor operating an ISO would be incapable of altering its intentions. Rather, the assessment adds urgency for diplomatic communications.

In brief, the ISO Communication Paradox suggests a formula:

- Communication Necessity $\propto$ Threat Level
- Communication Probability $\propto$ 1/Threat Level

The first equation says that communication necessity is proportional to Threat Level. As threat level increases, the necessity for communication attempts increases at the same rate. The second equation says that communication probability is proportional to 1/Threat Level. This presents an inverse relationship. As threat level increases, probability of communication success decreases, and if threat doubles, the probability is cut in half.

In simple terms, a higher threat equals a more urgent need to communicate, while a higher threat also equals less chance that communication will work. Thus, we have a decision matrix where maximum communication urgency coincides with minimum success probability, requiring protocols that operate under deteriorating diplomatic conditions.

## II.     Core Propositions and Assumptions Establishing a Need for SCPs

*Five Propositions*

Five propositions about the relationship between ISOs and communication serve as contextual background for understanding the approach to developing communication protocols. They are:

- **Proposition 1 (P1): Communication carries risk; therefore, ISO threat assessment must outweigh the threats posed by communication.**
  Human to human communication is complicated, and cross-linguistic and cross-cultural communication more so given translation complications as well as differences across "beliefs, values, norms, assumptions, traditions, and expectations" (Johnston, 2003). Communication with a technological ISO-6+ therefore carries inherent risks of miscommunication. However, silence in the presence of key threat indicators carries greater risks of unopposed hostile action. Communication should therefore be initiated when threat assessment algorithms indicate that diplomatic engagement offers superior outcomes to passive observation, even accounting for miscommunication risks.

  Taking a conflict resolution perspective, there are benefits to communicating since communications can prevent conflict and since communication enacted in the middle of an immediate crisis will compete with more constraints and may be deemed less useful, less significant, or as arriving too late (Wojcik, 2019; Touitou, 2020; "Crisis," 2022). Although conflict resolution literature has been developed for human-to-human communication, there is no current way to theorize human to alien communication. Nevertheless, the idea that it is easier to establish peace prior to conflict just as it is easier to convince an animal or a person to cease planning an attack than to stop them after they have initiated an attack is presumed as a rule.

- **Proposition 2 (P2): ISO size, speed, trajectory, technosignatures, and propulsion evaluations are minimum necessary requirements to start communication decisions.**
  A suspected active technology at ISO-6 is the lowest possible level of ISO to which communications should be considered. That is to say, an ISO with non-natural technosignatures or signs of being operational or having propulsion capabilities presents indicators of risk strong enough to debate communication attempts. Such factors must be evaluated within the context of an ISO's trajectory, size, and speed, but these represent the baseline knowledge required for deploying a tool like the Threat-Communication Viability Index, which is designed for ISO-6+.

- **Proposition 3 (P3): Universal constants and harmonic mathematical communications are presumed to span intelligent species.**
  Universal constants and harmonic mathematical communications speculatively represent the most effective foundation for interstellar diplomatic contact because they are reasonable candidates to transcend cultural and species-specific barriers that could otherwise derail communication attempts. Astronomer Alexander Zeitsev (2006), for example, argues that "magic" wavelength values, such as "6.72 cm = 21 cm / Pi, would be known to all technological civilizations as the ratio of two universal constants, one physical (the radio emission line of interstellar neutral hydrogen) and the other mathematical" (5).

  Mathematical constants such as $\pi$, e, and the golden ratio $\varphi$ exist as fundamental properties of the universe, meaning any technologically advanced civilization capable of interstellar travel would probably recognize these relationships regardless of sensory apparatus or cultural development. When such constants are made into harmonic communication patterns, such as converting the digits of $\pi$ into frequency ratios, the effect for humans is aesthetically pleasing; of course, there is no guarantee that what is deemed as "aesthetically pleasing" transcends species, but the resulting message would demonstrate intelligence, sophisticated mathematical understanding, and pattern recognition.

- **Proposition 4 (P4): Communications initiated from Earth may signal distress, weakness, or naivete, favoring communication only in conditions of existential threat.**
  Initiating any communication on our end without receiving a message first provides extraterrestrial intelligence with a basis for psychological assessments and the following possibilities for explaining the attempt: a) an intelligent curiosity, b) a call for help or distress, c) the communicator's superiority (unafraid to reach out or give a command), d) a mere belief in the communicator's own superiority, or e) a palpable naivete. Option "c" may be ruled out presuming, of course, that an advanced technological object that flew millions of miles across the universe can indeed observe Earth's current technological state. This is good news in the sense that communicating superiority is undesirable since it can be construed with communicating a threat, and threat mitigation is at the heart of SCPs. Overall, we should expect human communications to signal a, b, d, and/or e. In brief, communication is a form of exposure, which strengthens the argument for reserving communication for specific threat assessment circumstances.

- **Proposition 5 (P5): Response pattern analysis requires communication strategy reevaluation and continued threat assessment.**
  (a.) Silence as a non-response to communication attempts requires evaluation against other threat indicators and the remaining communication window;
  (b.) Indecipherable responses increase threat probability but may indicate communication attempts requiring additional interpretation efforts;
  (c.) Similar-manner responses to human communications may but will not necessarily enhance peaceful interaction probability since extraterrestrial civilizations with both superior technology and destructive intents would either see no need to deceive or choose to deceive to move closer to Earth or to test human psychology. Notably, however, since communication only commences under conditions of high threats when one judges that there is more to gain than to lose from communication, interacting offers the opportunity to show intelligence and share Earth's value.

*Nine Base Assumptions about ISO-6+ Objects That Inform Communication Protocols*

The following nine assumptions describe why communications would be worth initiating. They articulate speculative ideas about ISO capabilities, and they are admittedly anthropocentric "best guesses" describing technological ISOs. Assuming these are correct, however, they imply the difficulties of communication but also show what communication efforts might achieve.

1. Presumed Monitoring Capability: Engages in continuous intelligent observation
2. Presumed Detection Capability: An ISO-6+ can detect life on Earth
3. Presumed Communication Capability: Possesses signal detection and transmission capabilities
4. Presumed Maneuvering Capability: Has control over trajectory and velocity modifications
5. Presumed Decision Flexibility: Controllers may have willingness to alter actions
6. Presumed Cultural Unfamiliarity: Unlikely to understand human languages or cultural contexts
7. Presumed Peace Familiarity: Have peaceful ideas and rationales that can be appealed to
8. Presumed Mathematical Recognition: Able to identify mathematical conventions and constants
9. Presumed Intelligence Signaling: Mathematical harmony and predictability communicate intelligence and reasoning

### III. Threat-Communication Viability Index (TCVI) for SCPs

A Threat-Communication Viability Index (TCVI) aims to mathematically represent the ISO Information-Communication Paradox and the five core propositions and nine assumptions about ISOs, providing an algorithmic assessment for communication decisions. The TCVI intends to function as an aid to decision makers and is stated as follows:

**TCVI = (Loeb Level ÷ √Remaining Time) ÷ (Success Probability × Comm Window)**

The Threat-Communication Viability Index balances four key variables to produce a score. In the numerator, the Loeb Scale threat level (ranging from 1-10) is divided by the square root of remaining time until the ISO's closest approach, creating a "threat proximity factor" that grows as the object nears Earth. The square root dampening prevents the index from escalating too rapidly at long distances, allowing extended observation periods for distant objects while still capturing the accelerating urgency as approach time shortens. This numerator is then divided by two constraining factors. The first is the success probability, which decreases based on the number of hostile indicators detected, such as weapons signatures or trajectory changes. The second is the communication window, which is a mathematical representation of the realistic time needed to prepare, transmit, receive, and decode a message exchange assuming extreme difficulty, tense negotiation among stakeholders, and possibly the need for code-breaking.

Overall, the formula's architecture aims to reflect interstellar diplomacy under threat conditions. Stated simply, when an ISO is far away with few threatening characteristics, the denominator remains large (high success probability, adequate communication window) while the numerator stays small (low threat proximity), producing a low TCVI score that recommends continued observation. As the object approaches or displays more hostile features, the threat proximity factor increases while success probability decreases, driving the TCVI score upward and triggering escalating response protocols.

Notably, "Success Probability" in the formula represents direct threat variables, or hostile feature assessments that add urgency and specificity to the Loeb Scale.[3] In another register, "Success Probability" stands as a best guess at achieving communication outcomes given the confirmation of hostile features and the context of the communications. The measure does not assume 100% success but starts at 50% success as an ideal and entails a significant multiplicative reduction factor for every additional Loeb Level and confirmed threat variable.[4]

---

[3] We use the 9 threat variables, as outlined in this paper; the details of those variables can be adjusted, but if variables are added or subtracted, then mathematical consistency for the TCVI should be achieved to ensure the TCVI scale continues to offer viable recommendations fitting to the core propositions and base assumptions.
[4] 50% is an estimate referencing a best-case scenario for communications with an alien species. The number assumes only half of all communication attempts will deliver the intended message. As a base point and estimate, 50% reflects neither an overly optimistic nor pessimistic assessment in the face of no knowledge of alien intelligence and is chosen because an advanced intelligence may just as well comprehend all human languages quickly as use a totally foreign form of communication with no bridges to human forms at all.

The TCVI score aids decision makers and thus has levels of recommendations. It moves from recommending observation to recommending debate about the value of communication efforts; subsequently, as time decreases and/or threats increase, it moves to recommending active communication and ultimately to recommending emergency communication and defensive measures. The square root dampening in the threat proximity calculation creates a manageable escalation curve rather than exponential panic, allowing decision-makers to progress through deliberative phases as discussion continues. This mathematical structure attempts to formalize the otherwise paralyzing question of when the risks of communication are outweighed by the risks of silence when facing a potentially hostile technological ISO.

*TCVI Example 1*

**Scenario:** ISO-8 on an unusual but non-impact trajectory, showing three hostile technological indicators.

- **Threat Level on the Loeb Scale (8):** In this first example, we use ISO-8, which means "Confirmed Technology of Extraterrestrial Origins."
- **Remaining Time (45 days):** We imagine 45 days before it passes at the nearest point.
- **Communication Window (5 days):** For this example, if we estimate 650 million miles per hour for signal transmission and a 45-day arrival time for an ISO travelling at 100,000 miles per hour, then the communication signal reaches the target in 10 minutes or so, but we add 10 more minutes for return signals plus several days factoring in signal preparation and interpretation teams on both sides. We add 5 days in this case, which provides space for negotiations about signal creation yet allows time for multiple signal exchanges or follow-up attempts at communication given the Remaining Time.

The table below offers signal exchange periods to use in the communication window portion of the TCVI equation. They have been arranged algorithmically, which is to say prepared to ensure the TCVI reflects appropriate urgency for nearby threatening objects while staying realistic about what it requires to coordinate signal preparation on the ground. Although not every signal will take the same amount of time to prepare, these times leave space for thought while also enabling the TCVI to function as a decision-making tool.

**Signal Preparation Times for the TCVI Equation**

**For ISO Loeb Level 6-8**

- ≤ 15 days until ISO arrival, 1 day prep added to signal arrival times
- 16-30 days, 3 days prep
- 31-60 days, 5 days prep
- 61-120 days, 7 days prep
- 121+ days, 9 days prep

**For ISO Loeb Level 9-10**

- ≤ 30 days until ISO arrival, 1 day prep added to signal arrival times
- 31-90 days, 3 days prep
- 91-120 days, 5 days prep
- 121+ days, 7 days prep

**Success Probability (0.07 or 7%):** We start at 50% odds that any communication will be understood, but in the example, the Loeb Scale is at 8, reducing the starting point to 40% as a 5% reduction is applied per level above ISO-6 to mathematically factor in the added difficulty of communicating diplomatically amid threats. That is to say, as threat goes up, communication success goes down in keeping with the Information-Communication Paradox. In this example, the object is emitting high-energy radiation, changing course toward Earth, and jamming our communications. These three red flags each entail an additional 11% reduction in success probability (since there are nine key threat indicators which together would equal 99%); added together, these drop the chances 33% more, down to 7% (.4 – 0.33 = .07).

**The math for such an example works as follows:** We first calculate the threat proximity factor by taking the square root of remaining time: $\sqrt{45} = 6.708$. We then divide 1 by this value to get the threat proximity: **1 ÷ 6.708 = 0.1491**. Next, we multiply the Loeb level by this threat proximity factor: **(8 × 0.1491 = 1.193)**. We then multiply the communication window by the success probability **(5 days × 0.07 = 0.35)**. Finally, we divide the numerator by the denominator: **(1.193 ÷ 0.35 = 3.41)**.

A TCVI score of 3.41 in the above example means "begin immediate communication." In contrast, if we imagine 9 days as the communication window and all else remains the same, we have **(9 days × 0.07 success probability = 0.63)**. The final calculation then equals **(1.193 ÷ 0.63 = 1.89)**, which indicates that the communication recommendation at that stage remains in the same but the additional time slightly reduces urgency.

**The TCVI recommendations are as follows:**

- **TCVI < 0.15:** Extended observation and intelligence gathering
- **TCVI 0.15-0.24:** Prepare communication systems and protocols
- **TCVI 0.25-1.50:** Debate the value of communication attempts and decide
- **TCVI 1.51-3.50:** Begin immediate communication attempts
- **TCVI > 3.50:** Emergency communication with defensive activation

We can consider further examples:

*TCVI Example 2*
**Scenario:** ISO-6 on unusual but non-impact trajectory, showing one hostile technological indicator (High-energy radiation emissions).

- Threat Level (Loeb Scale): 6
- Remaining Time: 90 days
- Communication Window: 7 days
- Success Probability Calculation:
    - Base Success Probability: 0.5 (50%)
    - Threat Variable: High-energy radiation emissions present (-0.11)
    - Final Success Probability: 0.5 - 0.11 = 0.39 (39%)

**TCVI Calculation:** We first calculate the threat proximity factor by taking the square root of remaining time: $\sqrt{90} = 9.487$. We then divide 1 by this value: **1 ÷ 9.487 = 0.1054**. Next, we multiply the Loeb level by this threat proximity: **(6 × 0.1054 = 0.632)**. We then multiply the communication window by success probability: **(7 × 0.39 = 2.73)**. Finally: **(0.632 ÷ 2.73 = 0.23)**, **TCVI = 0.23**

This result sits below the debate range, arguing for continued observation and preparation.

Once the communication window lowers, assuming no change to the Loeb scale or threat indicators, the TCVI in this case will breach .25 but likely not breach 1.50. Notably, however, an increase in the level of the Loeb Scale would force the TCVI to breach 1.50 and require communication attempts.

*TCVI Example 3*

**Scenario:** ISO-7 on unusual but non-impact trajectory, showing one hostile technological indicator (sudden detection reduction).

- Threat Level (Loeb Scale): 7
- Remaining Time: 150 days
- Communication Window: 10 days (9 prep days plus transmission times)
- Success Probability Calculation:
    - Base Success Probability: 0.45 (45% - slightly lower due to Level 7)
    - Threat Variable: Detection capabilities seemed reduced (-0.11)
    - Final Success Probability: 0.45 - 0.11 = 0.34 (34%)

**TCVI Calculation:** Threat proximity = $1 \div \sqrt{150} = 0.0816$. Numerator = **($7 \times 0.0816 = 0.571$)**. Denominator = **($10 \times 0.34 = 3.4$)**. Final calculation: **($0.571 \div 3.4 = 0.17$), TCVI = 0.17**

This result does not require the international community to debate communication attempts, and the example allows us to extrapolate the fact that a Loeb Scale 7 without any threat variables using all other calculations here would not be deemed threatening enough to recommend risking communication attempts. Without threat indicators: threat proximity = **0.0816**, numerator = **($7 \times 0.0816 = 0.571$)**, denominator = **($10 \times 0.5 = 5$), TCVI = 0.11**.

Further, we can extrapolate that adding another threat variable to Example 3 would escalate urgency. With two threats in tow, success probability drops to 0.23: numerator = **0.571**, denominator = **($10 \times 0.23 = 2.3$), TCVI = 0.25**. Thus, debate must commence. In brief, the TCVI balances the hostile factors that appear as oddities with and against the non-impact trajectory, so the calculation represents the idea that communication should be considered even in non-impact situations given the characteristics of the object and potential unpredictability of threat that could emerge from such a compiling of characteristics.

*TCVI Example 4*

**Scenario:** Massive technological ISO on confirmed global impact trajectory showing none of the 9 additional hostile indicators; the trajectory itself is the primary issue.

- Threat Level (Loeb Scale): 10
- Remaining Time: 30 days
- Communication Window: 1 day (timeline from signal preparation times)
- Success Probability Calculation:
    - Base Success Probability: 0.3 (30% - low due to Level 10 crisis)
    - Threat Variables: None present beyond the collision trajectory itself
    - Final Success Probability: 0.3 (no additional reductions)

**TCVI Calculation:** Threat proximity = $1 \div \sqrt{30} = 0.1826$. Numerator = $(10 \times 0.1826 = 1.826)$. Denominator = $(1 \times 0.3 = 0.3)$. Final calculation: $(1.826 \div 0.3 = 6.09)$, **TCVI = 6.09**

This is an interesting case because an ISO at Loeb Level 10 represents a total existential threat. Accordingly, the communication window uses condensed timelines to accommodate the rule that existential threats require faster responses and earlier communications. In this case, the TCVI is at 6.09, well above the emergency threshold, and communication is absolutely necessary and pursued as a last resort.

*TCVI Example 5*

Thus far, all examples have explored ISOs approaching within 30-150 days. What if an ISO is detected early, and we have one year to study it?

This final example calculates the TCVI for such a scenario and demonstrates a key limitation of the formula: at extended distances beyond approximately 300 days, the TCVI scores compress into a narrow range where decision point thresholds become less distinguishable. The formula works optimally for objects within a 300-day arrival period because the square root dampening creates meaningful separation between threat levels and decision thresholds at closer ranges, but this separation diminishes at longer timescales.

**Scenario:** Early detection of an ISO of undetermined size having suspicious technological features and a likely impact trajectory; yet, the object is very far away.

- Threat Level (Loeb Scale): 9
- Remaining Time: 365 days
- Communication Window: 9 days
- Success Probability Calculation:
    - Base Success Probability: 0.35 (35% - low due to Level 9 crisis)
    - Threat Variables: None present beyond the collision trajectory itself
    - Final Success Probability: 0.35 (no additional reductions)

**TCVI Calculation:** Threat proximity = **1 ÷ √365 = 0.0523**. Numerator = **(9 × 0.0523 = 0.471)**. Denominator = **(9 × 0.35 = 3.15)**. Final calculation: **(0.471 ÷ 3.15 = 0.15)**, **TCVI = 0.15**

This calculation produces a TCVI at the boundary between observation and preparation, which seems inappropriately low for an ISO-9 with impact trajectory. The issue becomes clearer when calculating more threat scenarios at this distance. For example, an ISO-6 at 365 days with no threats yields: threat proximity = **0.0523**, numerator = **(6 × 0.0523 = 0.314)**, denominator = **(9 × 0.5 = 4.5)**, **TCVI = 0.07**. And an ISO-9 at 365 days (TCVI = 0.15) is only marginally higher than an ISO-6 (TCVI = 0.07), despite representing a vastly more dangerous scenario.

In brief, beyond 300 days, the square root dampening creates such small threat proximity values that even dramatic differences in Loeb levels or threat indicators produce minimal TCVI separation. This compression means the formula cannot effectively distinguish between scenarios that should trigger different responses. For ISOs detected beyond 300 days, alternative decision frameworks could provide more useful guidance than the TCVI.

*On TCVI Recommendations*

When to communicate and when not to communicate is a complex question that can only be determined on a case-by-case basis with relevant experts and the aid of the international community. The above calculations and the logic informing them is offered as one method to facilitate meaningful discussions. It is not offered as a definitive method, and the method is useful only when ISO arrival time is 300 days or less.

Overall, the TCVI ranges and recommendations suggest that when something dangerous is approaching and time is running out, one should try to communicate, even if the effort probably will fail because doing nothing guarantees the worst outcome.

### IV. Strategic Communication Protocol (SCP) Framework

The following protocols offer one set of communication approaches and recommendations designed to be used in accordance with the TCVI. That noted, this researcher encourages others to also create communication strategies fitting to the TCVI.

**TCVI under .24: Strategic Intelligence Gathering**
- Communication Approach: Maintain observational silence while conducting comprehensive threat assessment

**TCVI .25-1.50: Debate Phase - Consider the relative benefits of silence versus pivoting to the next protocol.**

**TCVI 1.51-3.5: Peaceful People Protocol**
- Communication Approach: De-escalation diplomacy with parallel defensive preparation
- Primary Protocol: Emergency diplomatic communication and quiet defense measures
- Initial Content Suggestion: Mathematical constants ($\pi$, $e$, $\varphi$), which could be followed by constants converted to harmonic frequency ratios
    - Example: $\pi$ digits 3,1,4,1,5,9 → frequency ratios 3:2, 1:1, 4:3, 1:1, 5:4, 9:8
- Secondary Content Suggestion: Golden ratio sequences, aesthetically pleasing musical ratios alongside of the object's trajectory: consider mathematical representations of the object's trajectory. However, if that trajectory intersects with Earth, pivot to 3.5+ protocol or communicate only slowing of that trajectory coupled with harmony signals to avoid accidentally communicating hostility.

- - Critical Constraint: Avoid signaling trajectories intersecting with Earth (even if true) or any chaos patterns to prevent misinterpretation. If no response, pivot to the 3.5+ protocol.[5]
- Timing: Measured, spaced and regular intervals to demonstrate reliability and calmness
- Method: Maximum transmission across all frequencies with measured timing
- Parallel Actions: Begin discrete defensive preparations including interceptors and probes[6]
- Decision Point: If no responses are received, continue with the protocol. If like-manner positive or harmonic messages are received, proceed to Elevated Protocol (below). However, if responses to this protocol are chaotic or aggressive, pivot to TCVI 3.5+ protocol or choose strategic silence.

**Elevated Protocol, Prepared for a Response**
- Communication Approach: Positive echo diplomacy with threat monitoring
- Primary Protocol: Systematic mathematical communication with readiness for echo-back communications.
- Initial Content Suggestion: Repeat any message received with a harmonic signal before and after; if in doubt about message content, send a new message that transmits a mathematical constant, or pivot to a more specific message such as the "slow trajectory" messaging. Allow time for internal debate on message type given the ISO characteristics; see TVCI formula recommendations.
- Channel Management: Return messages only on received frequencies, cease other channels to confirm active communication.
- Decision Point: If messages are unclear but harmonic in appearance, repeat protocol and see what is sent in return while allowing time for deciphering. If messages are chaotic or aggressive, proceed to TCVI 3.5+ protocol, or choose strategic silence.

**TCVI 3.5+: Existential Threat Protocol**
- Communication Approach: Emergency de-escalation with full defensive activation

---

[5] The message of "slow down" is preferred to "change course" since a command of slowing does not entail the same fear and may be able to communicate a desire for discussion in a way that "go away" does not.
[6] Abraham Loeb argues for developing interceptors and probes. See: Loeb, A. (2025). "Should we message 3I/Atlas?" https://avi-loeb.medium.com/should-we-message-3i-atlas-43455e323f69

- Primary Protocol: All-frequency emergency broadcast
- Initial Content Suggestion: Clear ISO trajectory recognition signals with chaos patterns linked to Earth impact, followed by harmony patterns linked to trajectory changes; here we adopt a more direct set of communications given the existential nature of the threat.
- Method: Continuous broadcast across all frequencies
- Parallel Actions: Make visible defensive preparations including the use of interceptors and probes to capture closer images or directly engage the ISO in space.
- Decision Point: If slowing or trajectory changes are observed, immediately switch to harmony-only signals. Debate the idea of disengaging probes designed for interception.
- Acknowledgment: Communication may be futile at this level

**Adaptive Response Guidelines for SCPs**

Communication Success Indicators:
- Structured signal complexity in responses showing intelligence
- Harmonic or aesthetic reply patterns
- Mathematical pattern recognition in responses or echo-back signals
- Trajectory modifications following communication attempts
- Reduced threat indicators such that the TCVI lowers

Communication Failure Indicators:
- Continued approach with threat indicators
- Increased aggressive behavior following communication
- Random or chaotic response patterns
- Active jamming of communication attempts
- Escalation of threat indicators such that TCVI increases

## V. Limitations

Several limitations constrain the applicability and reliability of SCPs and the TCVI.

1. **Temporal Constraints Limit Deployment**

The TCVI formula is calibrated for ISOs approaching Earth within a 300-day window. Beyond this timeframe, the mathematical relationships within the formula break down and produce unreliable recommendations. Long-duration scenarios require different decision-making tools.

2. **Equal Weighting of Key Threat Indicators May Not Reflect Those Indicators**

The nine key threat indicators identified for demonstration of this framework are currently treated as mathematically equivalent; each contribute an 11% reduction to the success probability in the TCVI formula. This equal weighting does not reflect the intuitive understanding that some, such as directed energy beams targeting Earth, represent more severe threats than others, such as gravitational anomalies. However, the adoption of equal weighting is preferred given the profound uncertainty surrounding ISO capabilities.

Differential weighting only adds speculation to speculation. Without empirical data about extraterrestrial technology or hostile intent, weighting threats could potentially skew the conclusions. For example, observing a minor anomaly such as a temporal displacement effect may represent an extremely sophisticated and dangerous capability compared to many others even though it could be given comparatively less weight than other capabilities. Similarly, the absence of certain threats might reflect concealment. Overall, equal weighting acknowledges epistemic limitations and avoids false confidence in assessing ISO threat hierarchies.

3. **Multiple ISOs Showcase a Scenario Gap**

While the framework identifies "multiple coordinated objects" as a key threat indicator within the success probability calculation, the SCPs do not provide guidance for multi-ISO scenarios. The protocols assume communication with a single object or a single coordinated system, but offer no decision-making framework for situations involving multiple independent ISOs with potentially different trajectories, threat levels, or communication patterns.

Multi-ISO encounters introduce complex questions that exceed the current framework's scope: *Should we communicate with all or prioritize certain targets? How should conflicting signals*

*from different ISOs be interpreted? What if one ISO exhibits peaceful patterns while another shows hostile indicators?* These vexing questions require strategic communication planning. Nevertheless, the TCVI formula accounts for multi-ISO scenarios and the SCPs could still be applied as a starting point, recommending communication toward those ISOs that appear to be in control or primary. However, here we quickly see the framework's limitation as it regards multi-actor interstellar diplomacy. Thus, a future development should incorporate multi-ISO decision matrices.

4.  **Foundational Assumptions Risk Fundamental Error**

The entire SCP framework rests upon five core propositions and nine base assumptions about ISO capabilities and the nature of intelligence itself. If any single proposition or assumption proves incorrect, the protocols may fail or, worse, harm humanity's prospects for peaceful resolution.

For example, the framework assumes ISOs can detect life on Earth during their flyby, maintain intelligent monitoring at all times, possess communication capabilities, control their trajectories, and have decision-making flexibility. If an ISO is instead an automated probe with no capacity for course alteration or an artifact from a long-extinct civilization operating on predetermined programming, communication attempts are futile. More troublingly, if an ISO deliberately conceals its capabilities while monitoring human communication attempts, our protocols most certainly provide valuable intelligence about human psychology and technological limitations. Thus, diplomatic efforts risk exacerbating the odds of destruction.

That noted, the TCVI and SCPs together recommend reserving communication attempts to cases where the threat is evident and high. Indeed, situational awareness suggests that Earth and humanity sit in a strategically weak and passive position when confronting technological ISOs moving toward Earth at intense speeds, and consequently, inaction in that case is not evaluated as the best option. Consequently, we must discover some way to agree on decisions. The TCVI and SCPs aim to facilitate that.

**5. The Silence Alternative is Disregarded**

The framework acknowledges but does not adopt the idea that strategic silence represents the safest option even when facing apparent threats. If an ISO has not yet detected Earth as a technological civilization, communication only broadcasts our presence and location. If an ISO is conducting passive observation, remaining silent might allow humanity to be categorized as non-threatening. In addition, the decision to communicate involves profound uncertainty about extraterrestrial psychology, sociology, and strategic reasoning. Even so, the SCP framework chooses communication as the primary response to threat; perhaps problematically, this choice reflects human conflict resolution paradigms and may not apply to extra-terrestrial entities.

Although true, the SCPs adopt the idea that communication represents humanity's best attempt given the assumption that ISOs are aware of our presence and technological state since we broadcast radio signals and since Earth is surrounded by satellites and space junk. In other words, we communicate much of what we would fear communicating already and should, therefore, structure diplomatic engagement strategically. In addition, if an ISO is an ancient artifact on an unalterable trajectory, there is no harm done by communication attempts. In another register, of course, SCPs embody optimistic assumptions and function to manage uncertainty; however, this approach is preferable to paralysis in the face of existential threat.

## VI. Conclusion

The SCP framework operates within the constraints of the TCVI calculation and acknowledges that attempts at communicating with ISOs remain subject to the underlying Information-Communication Paradox. Nevertheless, the SCPs seek interstellar diplomatic engagement, adapting conflict resolution principles for first contact with a technological ISO. While the framework cannot guarantee successful outcomes, it provides structured approaches and seeks to maximize peaceful resolution probabilities.

The ultimate success of these protocols depends not only on their internal logic and structure, but on rapid international coordination, technological capability deployment, and the fundamental assumption that intelligence, whether human or non-human, can recognize and appreciate the universal language of mathematics, harmony, and peaceful intent.

In scenarios where communication fails and existential threats materialize, these protocols will serve as humanity's final diplomatic gesture. They are a testament to the human capacity for peaceful resolution, and they represent our hope that intelligence, wherever it emerges in the universe, shares some capacity for mercy, curiosity, and the preservation of life.

## References

Crisis Communication: A Behavioral Approach. *Government Communication Service.* https://www.communications.gov.uk/wp-content/uploads/2024/08/Behavioural-science-guide-to-Crisis-Communications.pdf

Eldadi, O., Tenenbaum, G., and Loeb, A. (2025). The Loeb Scale: Astronomical Classification of Interstellar Objects. https://arxiv.org/pdf/2508.09167

Johnston, C.B. (2003). Cross-Cultural communication. In D. Johnston, *Encyclopedia of International Media and Communications* (1st ed.). Elsevier Science & Technology. https://search.credoreference.com/articles/Qm9va0FydGljbGU6OTA2ODk1?aid=103034

Loeb, A. (2025). "Should we message 3I/Atlas?" https://avi-loeb.medium.com/should-we-message-3i-atlas-43455e323f69

Wojcik A. (2019). The Importance of Communication Before and During a Public Health Emergency. *Dela J Public Health* 5(4): 28-30.

Zaitsev, A. (2006). Messaging to Extra-Terrestrial Intelligence. https://arxiv.org/pdf/physics/0610031